# Lattice dynamics of the high temperature shape memory alloy Nb-Ru

S. M. Shapiro[1], G. Xu[1], G. Gu[1], J. Gardner[1,2], R. W. Fonda[3]

[1]*Brookhaven National Laboratory*
*Condensed Matter Physics/Materials Science Department*
*Upton, NY 11973*

[2]*NIST Center for Neutron Research*
*100 Bureau Drive, MS8562*
*Gaithersburg, MD 20899-8562*

[3]*Materials Science and Technology Division*
*Naval Research Laboratory Code 6356*
*Washington, DC 20375*

## Abstract

Nb-Ru is a high temperature shape memory alloy that undergoes a Martensitic transformation from a parent cubic β-phase into a tetragonal β' phase at $T_M \sim 900°C$. Measurements of the phonon dispersion curves on a single crystal show that the [110]-$TA_2$ phonon branch, corresponding in the q=0 limit to the elastic constant $C'=1/2(C_{11}-C_{12})$ has an anomalous temperature dependence. Nearly the entire branch softens with decreasing temperature as $T_M$ is approached. The temperature dependence of the low-q phonon energies suggests that the elastic constants would approach 0 as T approaches $T_M$, indicating a second order transition. No additional lattice modulation is observed in the cubic phase.

## Introduction

Shape memory materials (SMM) have the remarkable property of returning to their original shape after deformation, either by heating them to a temperature above the martensitic transformation (MT) temperature, or by unloading them when a stress-induced transformation has occurred[1]. A handful of metallic alloys exhibit this property. The most well-known and commercially developed material is the Ni-Ti alloy that was discovered in 1963 by Buehler et al. at the US Naval Ordnance Laboratory[2]. (This alloy was dubbed Nitinol as an acronym for <u>Ni</u>Ti <u>N</u>aval <u>O</u>rdnance <u>L</u>aboratory!). The applications to technology are limited only by an engineer's inventiveness. Its uses range from the medical (stents) to the mechanical (automated window openings) to the in-between: artificial sphincter muscles. Nitinol, and other commercially important shape memory alloys such as Cu-Zn-Al and Cu-Al-Ni, are useful up to temperatures of about 200°C. There is a technological need to develop new SMM for use at high temperatures that could be applied to automobile and rocket engines, gas turbines and nuclear reactor environments. Only a few alloys exhibit Martensitic transformation and shape memory properties at elevated temperatures of > 500°C. These include the Ti alloys, TiPt and TiPd[3] along with Ru alloys of NbTu[4,5] and TaRu[5,6]. Surprisingly, the studies of these are somewhat limited, most probably due to the difficulties of making the standard type of bulk measurements such as resistivity, magnetic susceptibility, etc. at these very high temperatures.

All SMM exhibit martensitic transformations (MT), which are at the heart of the shape memory property. These extensively studied transitions are always first order and transverse atomic displacements play a major role[7,8]. Precursor effects are usually seen, but their strength varies amongst the different systems. In some cases the MT is preceded by a pre-martensitic phase, in which a modulation of the lattice is superimposed upon the high temperature parent phase structure. The role of phonons as precursors to the martensitic transformation has also been extensively studied in many systems[9]. The majority of the studies show some phonon softening, but it is most pronounced in systems that exhibit a pre-martensitic phase. This raises the question of whether the phonon softening is a precursor to the MT or the pre-martensitic phase.

In this paper we report the first studies on a single crystal of the equiatomic composition SMM, NbRu. The phase transitions in this compound have been studied for over 35 years[3,4,6,10,11]. Two structural transitions have been reported between room temperature and 1000°C. Above 900°C, in the β-phase, the crystal structure is the ordered cubic, CsCl-type ($Pm\bar{3}m$) with the Nb and Ru atoms occupying the corners and the center of the cube. In reality this has not been proven because the x-ray scattering lengths of Nb and Ru are so similar that the superlattice peak intensities, corresponding to the differences in scattering lengths, are not observable. It was inferred from the similarity of the transformation with that occurring in TaRu alloy[12]. Below 900°C, the crystal transforms to a tetragonal structure, called the β' phase. There is another transformation near 750°C to β"-phase, which is either orthorhombic or monoclinic. The shape memory effects are associated with the high temperature the β to β' transition[5], which is first order, but with a surprisingly small amount of hysteresis. Earlier experiments claimed there was no hysteresis and the transitions can be considered as continuous, which contradicts a fundamental first order aspect of martensitic transformations. The first order nature of the β →β' transition was reported in more recent experiments that claimed a coexistence of the β and β' phases over a 200°C temperature range[12]. The β'to β" transition was reported to be continuous, which is allowed by the symmetry of the two phases according to Landau theory.

We report on temperature dependent inelastic neutron scattering experiments in the β phase of single crystals of $Nb_{50}Ru_{50}$, specially grown for this experiment using a state-of-the-art image furnace. We measure the partial dispersion curves along two symmetry directions and determined the elastic constants of the cubic phase. The only branch that shows a dramatic temperature dependence is the [qq0]-$TA_2$ branch corresponding to propagation along the [110] direction with perpendicular displacements along the [1-10] direction. This corresponds, in the limit of q → 0, to the elastic constant $C'=1/2(C_{11}-C_{12})$. There is a general softening over nearly the entire branch as the MT is approached, but the branch never approaches zero. No modulation of the lattice was observed nor is there a particular q-value of the softening.

## Experimental

The single crystal of $Nb_{50}Ru_{50}$ used in this experiment was grown at Brookhaven National Laboratory. High purity Nb and Ru were melted by using an arc-melting furnace under Ar gas atmosphere and then the melt was cast into rods. These cast polycrystalline rods of NbRu, 12mm in diameter and 120 mm in length, were used as feed rods. A floating-zone furnace equipped with two ellipsoidal mirrors and two 3.5 kW halogen lamps was used to grow the single crystals. The growth chamber was first evacuated to $10^{-6}$ torr and then filled with high purity (99.999%) Ar gas to a pressure of 1 bar. The crystal growth was carried out under Ar flow. The feed rod and the seed rod were counter-rotated to achieve homogeneous heating of the floating zone and to promote the mixing of the elements in the zone, which is useful in maintaining steady growth. The shape of the solid-liquid interface of the seed rod was controlled by varying the rotation speed and traveling velocities between the feed rod and the seed rod, in order to grow large single crystals in the seed rod[13]. The growth velocity of the seed rod was 0.4mm/h and that of the feed rod was 0.1mm/h. The as-grown rod had a diameter of 6 mm and length of 210 mm. It was cut into 10 to 25 mm length sections.

The many domains present in the low temperature phase at room temperature complicates the alignment and verification of the quality of the crystal. However, as the crystal is heated into the cubic-β phase, the domains disappear and a single uniform peak is observed in the rocking curve with width <0.3°. The crystal was mounted in an A.S. Scientific Products high temperature furnace with a maximum temperature of 1600°C. Temperature regulation was better than 1°C at the elevated temperatures. Some measurements were made with the temperature being scanned at a heating and cooling rate of ~12°C/min. The chosen scattering plane was (001), which allowed for observation of [110]-$TA_2$ phonon branch. Most measurements were made in the (1,1,0) and (2,0,0) Brillouin zones.

All neutron experiments were performed on the BT9 thermal beam triple axis instrument at the NIST research reactor. Pyrolytic graphite (PG) was used as a monochromator, analyzer and filter placed after the analyzer. A fixed final energy of $E_f$=14.7 meV was chosen for the experiments. The collimation was either 40-40-40-80 or 40-20-20-40 depending upon the conflicting needs of intensity and resolution.

## Results

### *Phase transformations*

Niobium and ruthenium, coincidently, have nearly identical neutron scattering lengths[14]: $b_{Nb}$=7.054, $b_{Ru}$=7.03. The ratio of the intensities of the superlattice reflections of the CsCl lattice [ (h+k+l)=odd], whose intensity is proportional to $(b_{Nb}-b_{Ru})^2$, compared to the fundamental bcc peaks [(h+k+l)=even],whose intensity is proportional to $(b_{Nb}+b_{Ru})^2$ is 3 x $10^{-6}$. This very small ratio makes it difficult to observe the superlattice reflections. Also, since they have nearly the same number of electrons it is difficult to measure the superlattice peaks with x-rays and the earlier x-ray studies could not measure them[12]. However in our study of the parent phase we used two PG filters to eliminate any higher order neutrons and were able to observe a weak (1,0,0) Bragg peak intensity. This confirms the expected, though unproven, β-CsCl ($Pm\overline{3}m$) symmetry of the high temperature cubic phase.

The early studies[4] showed that two phase transformations take place as a function of temperature. The high temperature β phase is cubic and transforms on cooling into the reported tetragonal phase (β') near 900°C[10] and another transition occurs near 750°C into the β" phase which is either orthorhombic or monoclinic[10,12]. Our measurements on a single crystal could not establish the definitive structure of the different phases but we could monitor the lattice parameters of the various phases by performing mesh scans about the two accessible Bragg peaks, (1,1,0) and (2,0,0). Figure 1 shows the intensity contours for the (1,1,0) and (2,0,0) Bragg peaks at temperatures in the three different phases and Table I lists the lattice parameters determined from these intensity contours. The top row shows the peaks in the β-phase measured at 900°C at the appropriate positions using the cubic phase lattice parameter, a= 0.3176 nm. Upon cooling to 875°C into the β' phase field, the crystal splits up into domains as evidenced by the multiple spots observed in the second row. For the measurement about (2,0,0,) the change in the distance of the spots along the H direction corresponds to the change in the lattice parameters, whereas the change along the K direction corresponds to the relative orientation of the domains. The lattice parameters of the β' phase measured in this way are $a_t$=0.3106 nm and $c_t$=0.3307 nm, which agree well with the high temperature x-ray measurements[12]. The third row are measurements at 600°C, where the symmetry is reported to be orthorhombic[12] or monoclinic[15]. The lattice parameters determined in this phase are $a_o$=0.3014nm, $b_o$=0.3081nm, and $c_o$=0.3430nm. The values agree well with Ref. 12 and our results are more consistent with the reported orthorhombic structure for the β" phase rather than monoclinic.

Although these transitions are expected to be first order, the early studies[4] of the resistivity and magnetic susceptibility showed very little hysteresis. (<10°C), but a large transformation temperature range of ~ 200°C. A later study[15] also reported a small hysteresis, but a narrower transformation range of ~40°C. We monitor the transformation temperatures by setting the spectrometer at the angles associated with a Bragg peak of a particular phase and then scanning the temperature while monitoring the intensity. For example, as seen in Fig. 1, if we align the spectrometer for a maximum of the (1,1,0) or (2,0,0) cubic peak intensity and cool the crystal, the intensity will disappear at the transition temperature as the lattice parameter changes and the Bragg peak shifts away from the settings of the spectrometer. This is shown in Fig. 2a, measured on heating and cooling. The β to β' transition is clearly seen and almost no hysteresis is observed. For this spectrometer setting there is no evidence of the β' to β" phase transition. On the other hand, if we set the spectrometer for one of the domains in the β' phase and vary temperature through the various phases we observe features that appear in both the β' and β" phases but are absent in the β phase. This is shown in Fig. 2b. The β to β' transition is present with little hysteresis, but a peak appears near 800 °C on heating and a bump on cooling near 700°C, which suggests a much larger hysteresis for the β' to β" transition.

### *Lattice dynamics and elastic constants*

The primary purpose of this experiment is to investigate the phonon behavior of the parent phase of NbRu and to look for precursors to the martensitic transition. Since the crystal was oriented in the (HK0) scattering plane it was only possible to measure along the [100] and [110] symmetry directions. Figure 3 shows the dispersion curves for the TA modes and the partial dispersion of the LA modes measured at 1100°C. Table II lists the velocities of the LA and TA modes determined from the limiting slopes near the origin of reciprocal space. From these values the elastic constants, tabulated in Table III, are calculated using the formula:

$$C = \rho v^2 \qquad (1)$$

where $\rho$ is the density: 6.05 gms/cm$^3$ and v is the limiting sound velocity. The anomalous branch is the [110]-TA$_2$ branch with atomic displacements along the [-110] direction and the limiting slope corresponds to the elastic constant $C'=1/2(C_{11}-C_{12})$. The anisotropy factor $A=C_{44}/C'=4.8$ is large, compared to harmonic lattices where it is unity. However, it is small compared to other shape memory alloys. For example, NiAl[16] Ni$_2$MnGa[17], NiTi[18], AuCd[19] have values of A=23, 23, 2, and 14, respectively, for a given temperature and composition.. This branch also shows a slight curvature near the middle of the Brillouin zone, which as shown below is very temperature dependent.

### *Temperature dependent phonon branches:*

The only branch that shows any anomalous temperature dependence is the [110]-TA$_2$ transverse acoustic mode. Figure 4 shows the dispersion of this branch measured at 910°C, just above the transformation temperature, and 1550°C, the maximum limit of the temperature range of the furnace. At the higher temperature the dispersion is nearly a normal sinusoidal shape. As the temperature is lowered there is a softening of the branch nearly out to the zone boundary. This is shown more dramatically in Figure 5, which shows only the low-q portion of the dispersion curve measured at several temperatures between 910°C and 1550°C. A gradual softening with temperature is observed, but nowhere does the energy approach zero as the transformation temperature is approached. Also, there is no particular q-vector that shows a stronger anomaly as observed in other SMM such as NiAl[20] or Ni$_2$MnGa[21] alloys. These materials exhibit a pre-martensitic behavior which is a commensurate modulation of the cubic lattice. Figure 6 is a plot of the phonon energy squared vs temperature for several q-values near the zone center. These all extrapolate to zero at temperatures well below the transformation temperature, $T_M$ = 900°C, with the larger q-values extrapolating to a lower temperature. The temperature dependence of the TA phonon propagating along the [001] direction was also measured. This corresponds to the C$_{44}$ elastic constant in the limit of q~0. Only a weak softening (<10%) of the energy of the phonon measured at q=0.1 was observed over the temperature range of 950 – 1250°C.

## *Discussion*

We present the first neutron study of the structure and lattice dynamics of the high temperature shape memory alloy NbRu. We have confirmed that the structure in the parent phase was simple cubic CsCl ($Pm\overline{3}m$) by observing a superlattice reflection, (1,0,0), which was not seen in the previous x-ray experiments due to the similar atomic scattering factors of Nb and Ru.

We also monitored the transitions between the three phases upon cooling: β ⟶ β'⟶ β". The β to β' transition occurs at 900°C in agreement with earlier measurements[15]. The abrupt transition is over a very narrow temperature range (<10°C) and shows very little hysteresis. The transition to the β" phase occurs at ~800 °C on cooling and ~725°C on heating. The lattice parameters of these phases have been determined and are consistent with previous measurements. The structure of these phases cannot be determined in this experiment and will await future high temperature powder diffraction studies of this material. There is agreement that the β' phase is tetragonal, but there is debate about whether the β" is orthorhombic or monoclinic. Our measurements are more consistent with an orthorhombic structure.

The dispersion curves of four phonon branches were measured and the elastic constants of the cubic phase and the anisotropy factor A were determined. As in nearly all materials exhibiting martensitic transformations, the [110]-TA$_2$ branch has the lowest energy and the largest temperature dependence. This was predicted many years ago by Zener[22] who explained that an intrinsic instability exists in an open bcc structure with respect to the [110]-TA$_2$ shear, due to the cancellation between short range exchange forces and longer range electrostatic forces.

The temperature dependence of this branch shows a gradual softening on approaching the β ⟶ β' transformation temperature. The softening occurs over 80% of the Brillouin zone. There is no particular q-vector where the softening is more singular as observed in many SMM such as NiAl[20], NiTi[23], AuCd[24], Ni$_2$MnGa[21]. In these systems the cubic lattice exhibits a modulation as evidenced by the appearance of diffuse elastic scattering at a particular wavevector $q=q_0$. The [110]-TA$_2$ branch in these systems exhibits an anomaly at the same wavevector $q_0$ as elastic diffuse scattering. This diffuse scattering intensity grows and eventually becomes a Bragg peak of either the martensite phase or an intermediate phase, which can be viewed as cubic with a short wavelength modulation of the lattice. In NbRu, elastic scans along the [1-10] direction relative to the (1,1,0) Brillouin zone center showed no extra peaks as evidence of any modulation. Some diffuse elastic scattering appears very near the Bragg peak, which increases as the transition temperature is approached.

The softening observed is also distinct from what has been observed in the alkali metals[25]. In these systems, there is nearly no softening near the zone center, but the amount of softening increases as q increases towards the zone boundary. The maximum softening, however is only ~10%.

The behavior shown in Fig. 5 is reminiscent of what is observed in the invar alloy, $Fe_{1-x}Pd_x$, which undergoes an fcc-fct transition at temperatures that are composition dependent[26]. Here, a normal linear dispersion curve is observed at temperatures above the magnetic ordering temperature $T_C$=575K. As T is decreased below $T_C$ there is a gradual softening of the entire [110]-$TA_2$ branch out to the zone boundary. Since the softening commences below the Curie temperature, the temperature dependence arises from a coupling of the phonons to the magnetization. In NbRu, if the softening is due to coupling to other degrees of freedom it is unclear what they are. A most likely candidate is the phonon coupling to electronic degrees of freedom as occurs in many other materials exhibiting martensitic transitions and has been shown to be the driving force of the transition in several martensites[27]. First principle calculations of the electronic structure and electron phonon coupling in this material are highly desirable

Figure 6 shows a plot of $(\hbar\omega)^2$ vs T for several q-values. This plot is justified in the soft mode theory of structural phase transitions[28], which states that the soft mode frequency in the classical theory varies as:

$$\omega_{SM}^2 \sim (T - T_C) \qquad (2)$$

where $T_C$ is the transition temperature. In a continuous second order transition, the soft mode frequency goes to 0 at $T_C$. In a first order transition, the soft mode is finite at $T_C$, but extrapolates to 0 at $T_E$. The extrapolation to zero energy varies with q. Figure 7 shows the extrapolated temperature as a function of q for q=0.1, 0.15, and 0.2. This linear curve extrapolates to $T_E$ = 892°C for q=0, which is very close to the martensitic transtion temperature in this system. This implies both that the β to β' transition is a second order transition and that the elastic constant C', which is a measurement at q=0, would go to zero at the transition. No elastic constant measurements have been made in this material and these results suggest that they be performed and would reveal some very interesting behavior.

*Summary*

We have monitored the martensitic phase transformations and measured the anomalous phonon behavior in the high temperature shape memory material NbRu. The high temperature β phase is cubic CsCl structure ($Pm\overline{3}m$) and the transformation occurs at T~900°C with nearly no hysteresis. The lattice dynamics in the β-phase shows an anomalous temperature dependence of the [110]-TA$_2$ phonon branch. There is a gradual softening over nearly the entire Brillouin zone, but no particular q-value where the softening is the largest. The elastic scattering measurement along this direction does not reveal any diffuse peaks at particular q-values. Plots of the temperature dependence of the softening suggest a second order nature for this transition as well as an anomalous behavior of the elastic constants.

## Acknowledgements

The authors thank Lee Tanner, Peter Vorderwisch, K. Otsuka and Simon Moss for helpful discussions. The work at Brookhaven is supported by the Office of Science, U. S. Department of Energy, under Contract No. DE-AC02-98CH10886. We acknowledge the support of the National Institute of Standards and Technology, U.S. Department of Commerce, in providing the neutron research facilities used in this work. RWF acknowledges support from the Naval Research Laboratory under the auspices of the Office of Naval Research.

# Figure Captions:

Figure 1: Intensity contours of elastic scattering for spectrometer set for the β-phase (1,1,0) (left side) and (2,0,0) (right side). Top row is measured at 900°C in the β-phase. Second row is 875°C in the β'-phase. Third row is measured at 600°C in the β"-phase.

Figure 2: Temperature scan for Bragg peak intensities. Top: Spectrometer set for (1,1,0) cubic β–phase Bragg peak. Bottom: Spectrometer set for one (1,1,0) variant of β'-phase Bragg peak.

Figure 3: Partial phonon dispersion curves for Nb-Ru measured at 1100°C in the β–phase.

Figure 4: Temperature dependence of the [110]-TA$_2$ branch of NbRu measured at two temperatures in the β-phase.

Figure 5: The low q-portion of the [110]-TA$_2$ phonon branch of NbRu measured at several temperatures in the β–phase.

Figure 6: Square of the phonon energy vs temperature for three different q-values. The line is a linear fit to the data. The transformation temperature, $T_M$ is indicated on the figure.

Figure 7: The extrapolated temperature ($T_E$) determined in Fig. 6 as a function of q.

Table I
Comparison of lattice parameters of $Nb_xRu_{1-x}$

| Phase | | Present Results (nm) $Nb_{50}Ru_{50}$ | Previous measurements(nm)[12] $Nb_{44}Ru_{56}$ |
|---|---|---|---|
| β-phase (T=900K) | $a_c$ | .3176 | .3818 |
| β'-phase | $a_t$ | .3106 | .3110 |
| | $b_t$ | .3307 | .3332 |
| β''-phase | $a_o$ | .3014 | .2993 |
| | $b_o$ | .3081 | .3063* |
| | $c_o$ | .3430 | .3416* |

* These values are those from Ref. 12 divided by $2^{1/2}$ to reflect the different choice of a unit cell in the present experiment.

TABLE II
Velocity of sound ($10^4$ cm/sec)
(measured at 1100°C)

| Direction | LA | TA |
|---|---|---|
| [011] | 58.5 | 25.4 |
| [110] | 60.9 | 11.6 |

TABLE III
Elastic Constants
(Measured at 1100°C)

| Elastic Constant | $10^{11}$ dynes/cm$^2$ |
|---|---|
| $C_{11}$ | 20.7 |
| $C_{44}$ | 3.9 |
| $C'=1/2(C_{11}-C_{12})$ | 0.81 |
| $C_{12}$ | 19.1 |

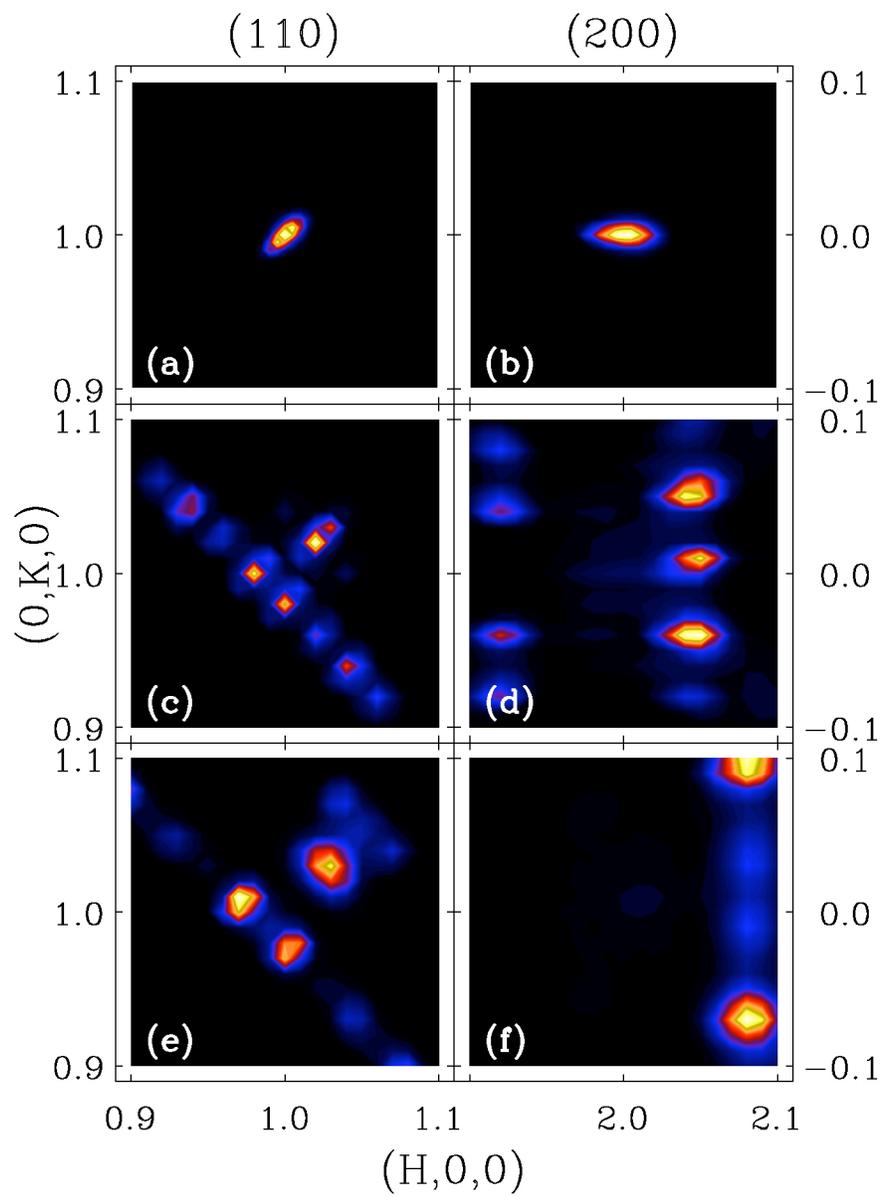

Figure 1

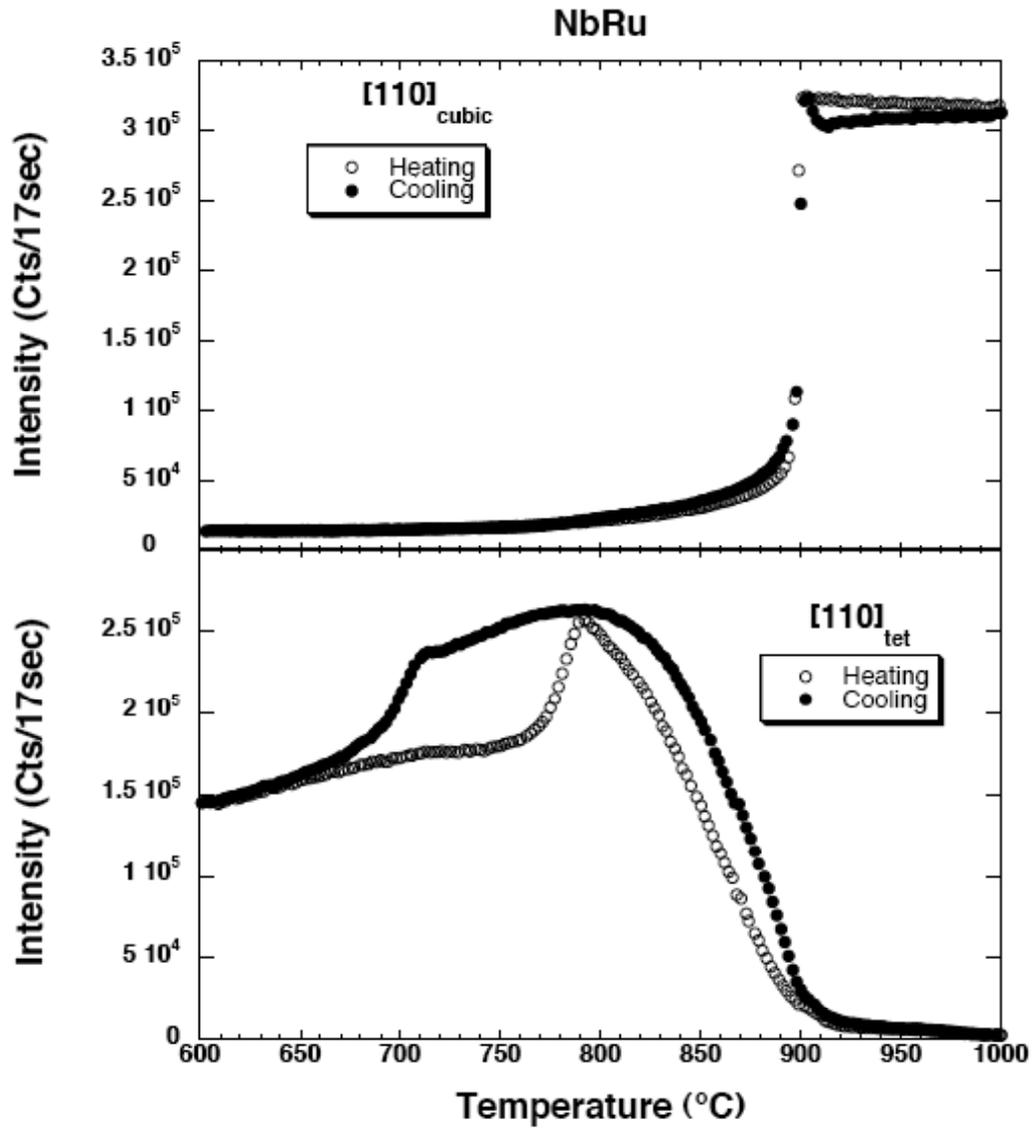

Figure 2.

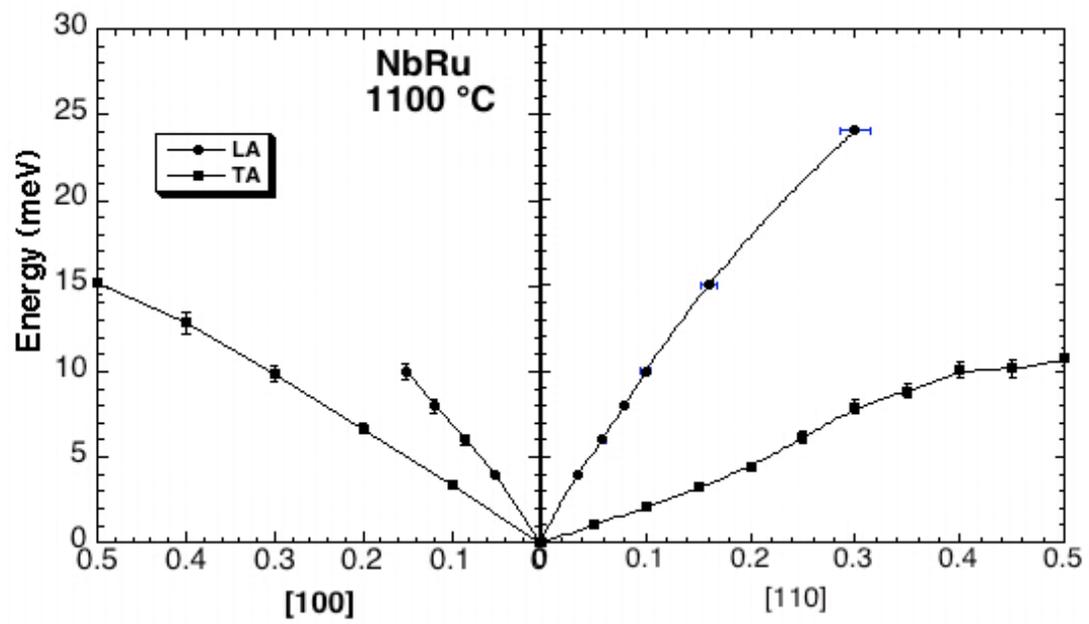

Figure 3

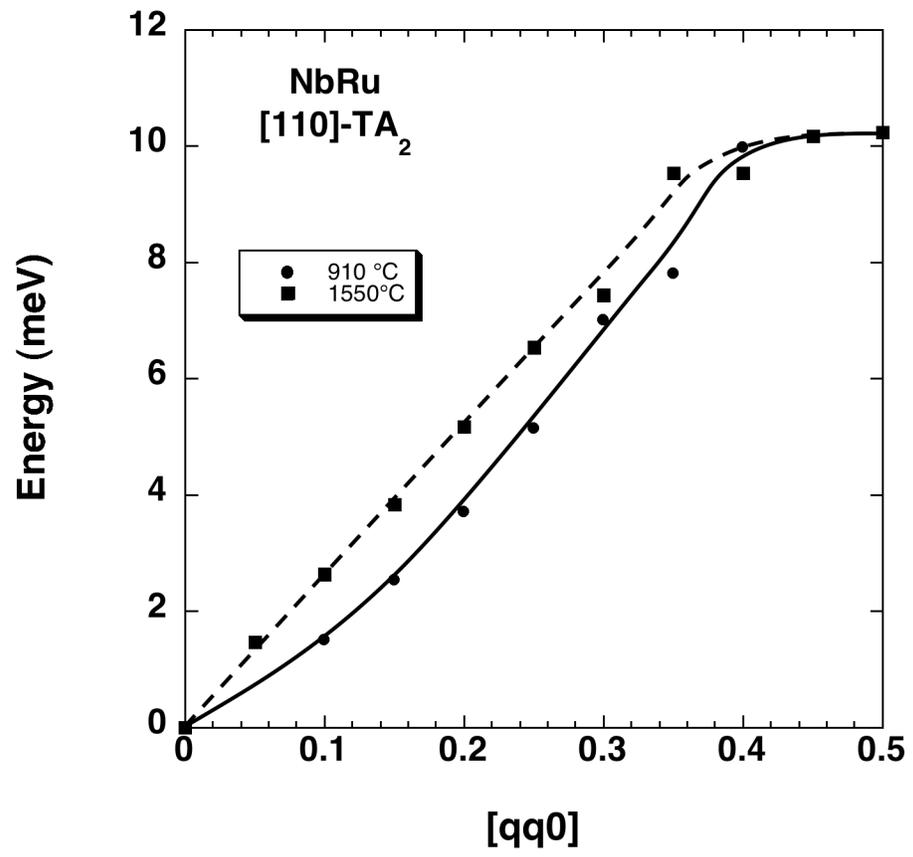

Figure 4

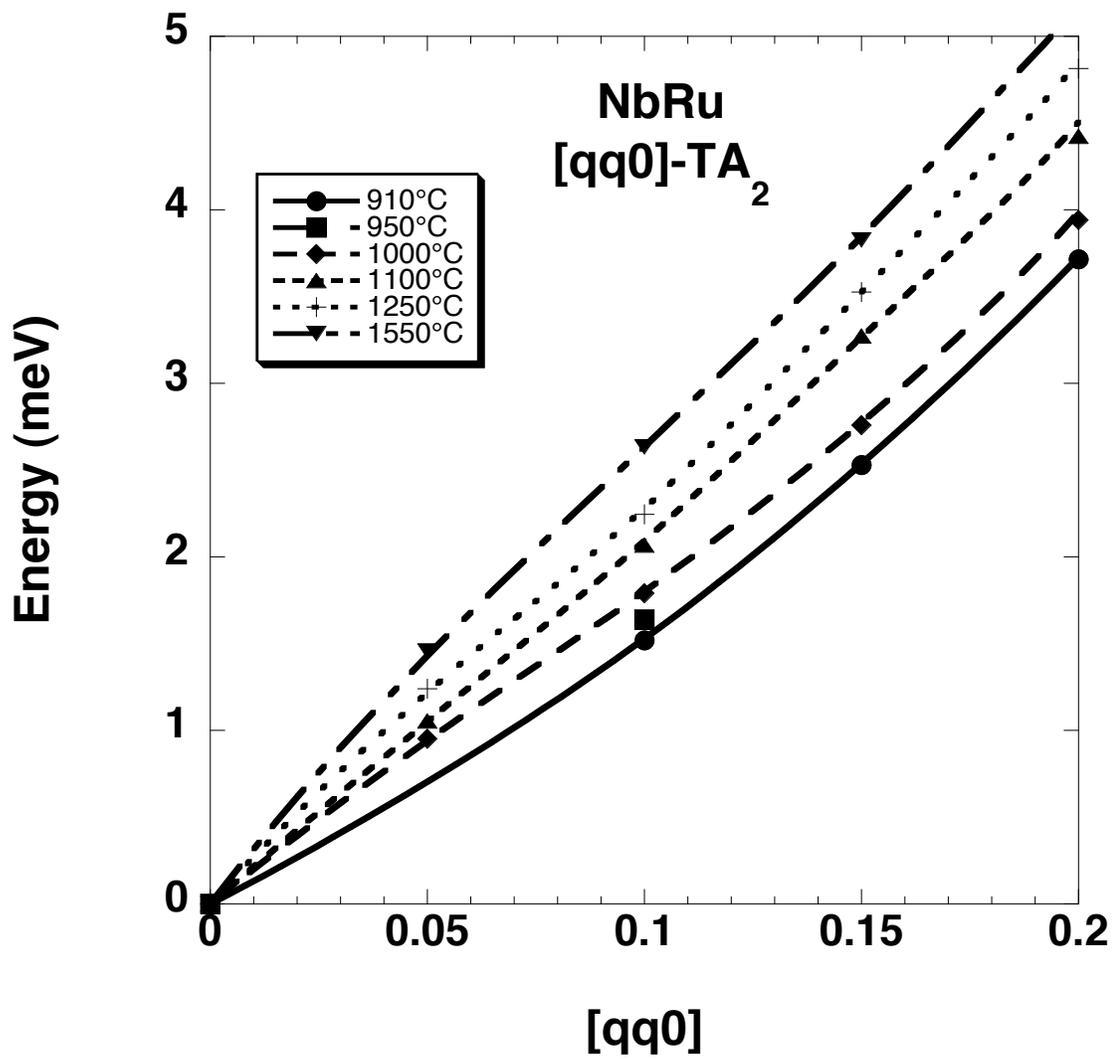

Figure 5

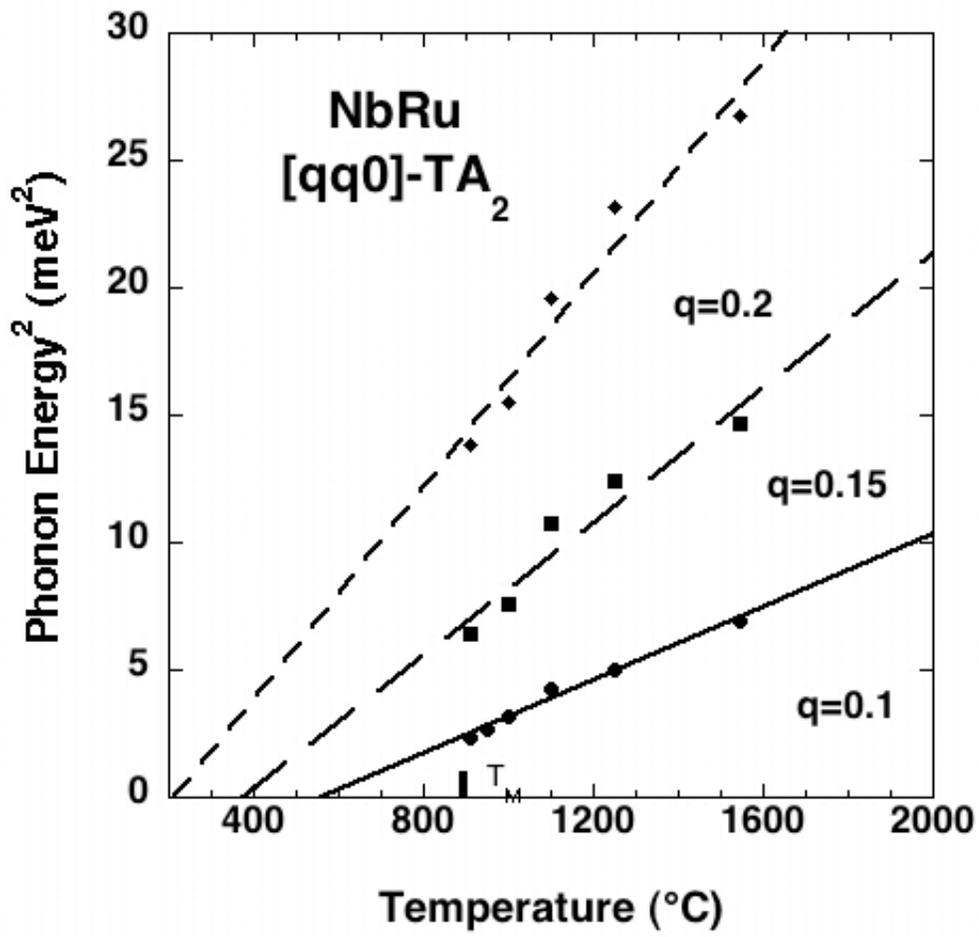

Figure 6.

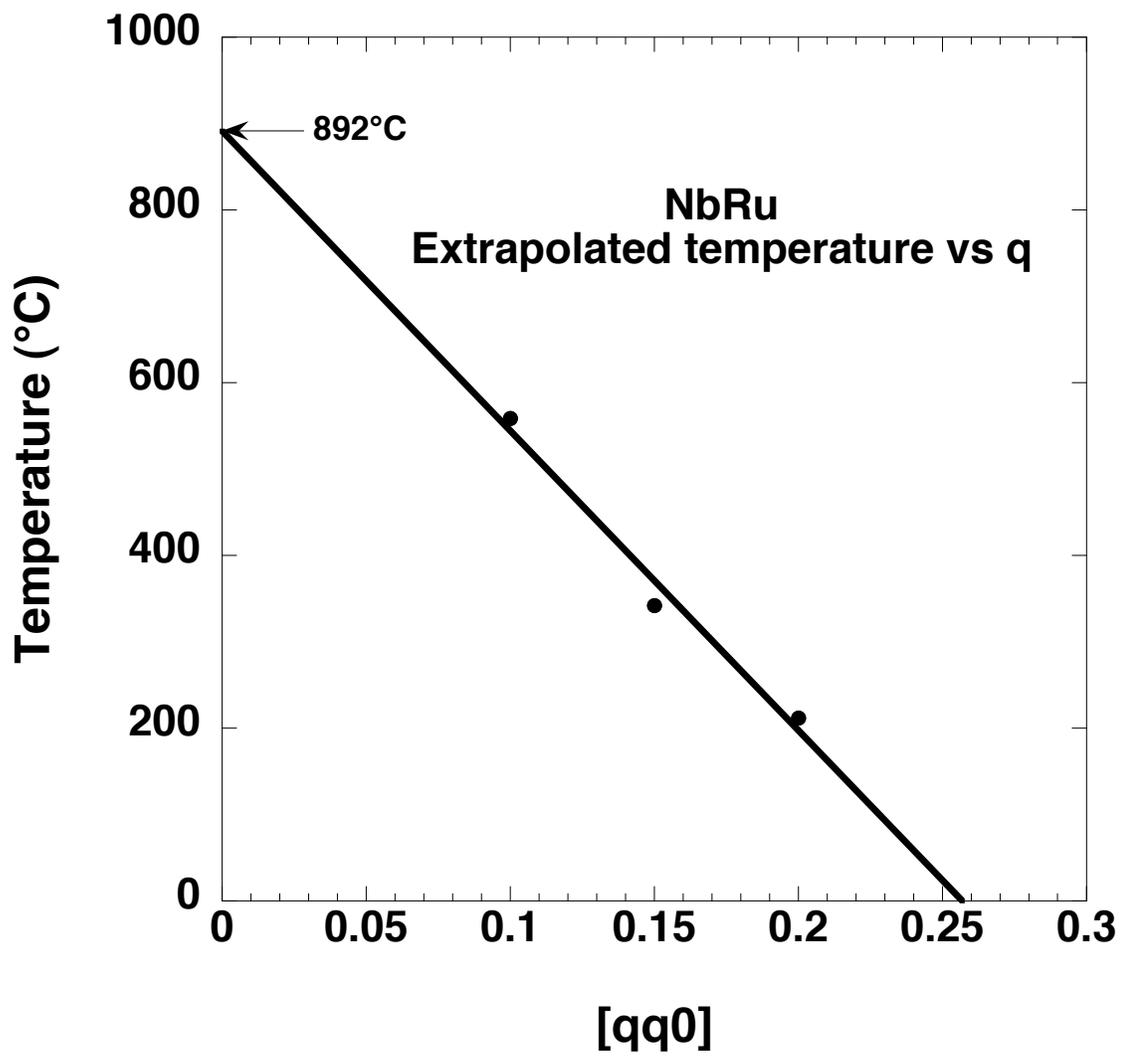

Figure 7.